\documentclass[authoryear,preprint,12pt,times,review]{elsarticle}


\usepackage{amssymb}
\usepackage{amsthm}
\usepackage{floatrow}

\usepackage{xcolor}
\usepackage{array}
\usepackage{bm}
\usepackage{bbm}
\usepackage{mathtools}
\usepackage{multirow}
\usepackage{natbib} 
\usepackage{url}
\usepackage{soul}
\usepackage{graphicx}
\usepackage[pdfstartview=XYZ,
bookmarks=true,
colorlinks=true,
linkcolor=blue,
urlcolor=blue,
citecolor=blue,
pdftex,
bookmarks=true,
linktocpage=true, 
hyperindex=true
]{hyperref}

\usepackage{orcidlink}

\newcommand{\ruiman}[1]{\textcolor{black}{#1}}

\newcommand\orcidicon[1]{\href{https://orcid.org/#1}{\mbox{\scalerel*{
\begin{tikzpicture}[yscale=-1,transform shape]
\pic{orcidlogo};
\end{tikzpicture}
}{|}}}}
\usepackage{hyperref}
\newcommand{\vect}[1]{\boldsymbol{\mathbf{#1}}}
\newcommand\norm[1]{\left\lVert#1\right\rVert}
\DeclarePairedDelimiter\abs{\lvert}{\rvert}%
 \journal{JRSS-A Frontiers in Data Integration Special Issue}
\makeatletter
\g@addto@macro\@floatboxreset\centering
\makeatother

\begin{document}

\begin{frontmatter}



\title{Spatial data fusion adjusting for preferential sampling using INLA and SPDE\tnoteref{label1}}

\author{Ruiman Zhong\,\orcidlink{0000-0002-4681-9199}\corref{cor1}\fnref{inst1,label2}}
\ead{ruiman.zhong@kaust.edu.sa}
\author{André Victor Ribeiro Amaral\orcidlink{0000-0003-3748-6801}\fnref{inst1}}
\author{Paula Moraga\orcidlink{0000-0001-5266-0201}\fnref{inst1}}
\cortext[cor1]{Corresponding author}
\fntext[label2]{Mail address: Thuwal,
           23955-6900, Makkah, Saudi Arabia}
\affiliation[inst1]{organization={Computer, Electrical and Mathematical Science and Engineering Division, King Abdullah University of Science and Technology (KAUST)},
            city={Thuwal},
            postcode={23955-6900}, 
            state={Makkah},
            country={Saudi Arabia}}

\begin{abstract}
Spatially misaligned data can be fused by using a Bayesian melding model that assumes that underlying all observations there is a spatially continuous Gaussian random field. This model can be employed, for instance, to forecast air pollution levels through the integration of point data from monitoring stations and areal data derived from satellite imagery. However, if the data presents preferential sampling, that is, if the observed point locations are not independent of the underlying spatial process, the inference obtained from models that ignore such a dependence structure may not be valid. In this paper, we present a Bayesian spatial model for the fusion of point and areal data that takes into account preferential sampling. Fast Bayesian inference is performed using the integrated nested Laplace approximation (INLA) and the stochastic partial differential equation (SPDE) approaches. The performance of the model is assessed using simulated data in a range of scenarios and sampling strategies that can appear in real settings. The model is also applied to predict air pollution in the USA.
\end{abstract}

\end{frontmatter}

\section{Introduction}
\label{s:intro}

Spatially misaligned data arise in a wide range of disciplines. For example, air pollution levels, defined as a continuous surface, may only be available as point-level data obtained from monitoring stations installed at specific locations or as areal-level data through satellite imagery. Spatial models that aim to integrate spatial data available at varying spatial resolutions are preferred over methods that use just one type of data---as they enable optimal utilization of all available information and enhance inferential outcomes.
Bayesian hierarchical models offer a viable solution for the combination of multiple spatially misaligned data.
For instance, \citet{fuentes2005model} introduced a Bayesian melding model that assumes the presence of a latent spatial process underlying all types of observations, enabling predictions by fusing all data types. \citet{Berrocal2010}, on the other hand, devised a linear regression model with spatially varying coefficients, employing point data as the response variable and areal data as covariates. 
\citet{Moraga2017} presented an implementation of the Bayesian melding model for data fusion using the integrated nested Laplace approximation (INLA) \citep{rue2009approximate} and a modification of the stochastic partial differential approach (SPDE) approach \citep{lindgren2011}, providing a computationally efficient alternative to existing Markov chain Monte Carlo (MCMC) methods.

When point data is considered, preferential sampling may occur if the spatial underlying process and sampling locations exhibit stochastic dependence \citep{diggle2010geostatistical, cecconi2016preferential}. This phenomenon has significant implications for spatial modeling, affecting both prediction accuracy and statistical inference \citep{diggle2010geostatistical, conn2017confronting, cappello2022adaptive}. In model-based geostatistcs, \citet{diggle2010geostatistical} account for preferential sampling by defining the observational model and the sampling scheme based on a shared latent spatial process, which is estimated using Monte Carlo methods. In this setting, \citet{dinsdale2018} developed numerical approximation methods for the likelihood function, resulting in more accurate estimation than original Monte Carlo approaches. 
In a Bayesian framework, \citet{pati2011bayesian} developed an approach for preferential sampling and studied its theoretical properties under improper priors. \ruiman{}
Later, \citet{krainski2018advanced} implemented the model proposed by \citet{diggle2010geostatistical} within the Bayesian hierarchical framework using the INLA-SPDE approach, which offers computational efficiency and enables the prediction at fine spatial surfaces. Most recently, \citet{shirota2022preferential} extended the preferential sampling model to the bivariate geostatistical setting. 

\ruiman{The impact of preferential sampling in sequentially collected data, such as molecular sequence,
has also gathered attention. 
In this setting, preferential sampling occurs when sampling frequency depends on some latent effect
\citep{karcher2016quantifying}, and it has been shown that ignoring this dependence leads to systematical bias.
\cite{parag2020jointly} proposed a model that allows the parameters to change across time intervals defined by certain change points. Later, \cite{cappello2022adaptive} proposed an adaptive preferential sampling model by accounting for this dependence structure through a latent random field.}

\begin{figure}
\centering
\caption[example1]{Example of spatially misaligned data with point data observed under preferential sampling. Left: continuous surface and sampled locations. Middle: point data with surface values at the sampled points.
Right: Areal data with aggregated values of the continuous surface at areas.}
\includegraphics[width = 1\textwidth]{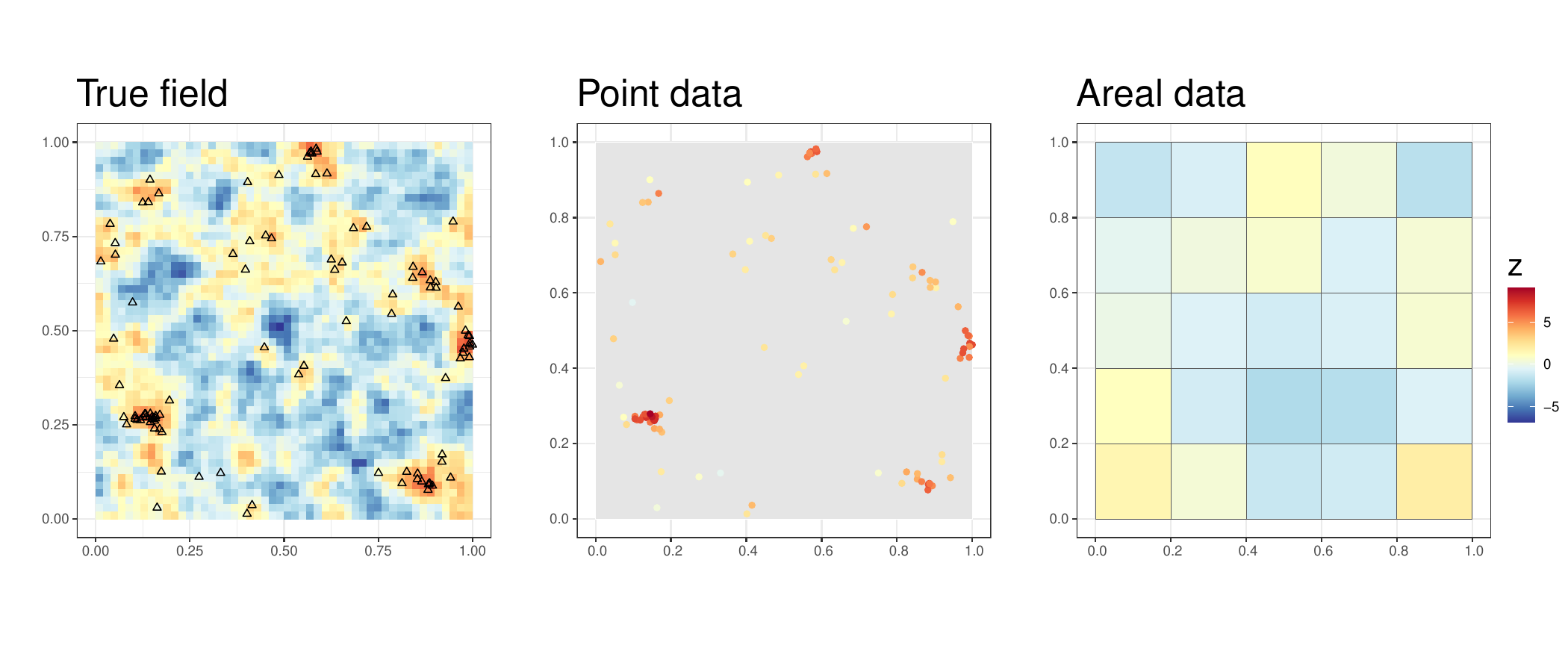}
\label{fig:sm_ps}
\end{figure}

In this paper, we propose a Bayesian hierarchical model to combine data available at several spatial scales adjusting for preferential sampling.
This model can be used in situations as the one represented in Figure \ref{fig:sm_ps}.
In that figure, the true latent spatial field is represented together with the sampling sites.
Note that the sites are concentrated in areas where the represented surface assumes high values.
The figure also shows areal data that arise from the true field aggregated into small areas. In situations like this, methods for combining  point- and areal-level data may also need to adjust for preferential sampling of point data to obtain valid inference. 

Motivated by \citet{fuentes2005model} and \citet{Moraga2017}, we specify a Bayesian hierarchical model to integrate data at different spatial resolutions by assuming a common spatial random field underlying all observations.
We also account for preferential sampling of point data by assuming a shared latent spatial random field for the intensity of the inhomogeneous Poisson point process with an additional parameter controlling the degree of preferential sampling, as introduced by \citet{diggle2010geostatistical}.
Additionally, we describe how to efficiently fit the model using the INLA and SPDE approach \citep{rue2009approximate,lindgren2011}. To the best of our knowledge, this study represents the first attempt to fuse spatial data while taking preferential sampling into account.
\ruiman{The implementation of our model is available at \url{https://github.com/RuimanZhong/PSmelding}.}

The remainder of this paper is structured as follows.
In Section \ref{s:sec2}, we present the new model to combine spatially misaligned data under preferential sampling.
Section \ref{sec:INLASPDE} gives details about the implementation of our method using the INLA and SPDE approach.
Next, we conduct a simulation study to assess the performance of the proposed model in comparison with two alternative models that do not combine data or not take into account preferential sampling.
In Section \ref{s:applicaton}, the proposed model is used to predict air pollution data in the United States. Lastly, our findings and future work are discussed.

\section{Spatial data fusion under preferential sampling} \label{s:sec2}

\subsection{Integration of spatially misaligned data}

Spatial misalignment occurs when measurements of the variable of interest are available at different spatial resolutions.
In these situations, a modeling approach that combines all these data sets may yield better inferences.
\citet{fuentes2005model}
proposed a Bayesian melding model to integrate point- and areal-level data by assuming a common spatial random field, say $\phi(\cdot)$, underlying all observations.
Specifically, if $Y(\vect{s}_i)$ represents data observed at locations $\vect{s}_i$, and $Y(B_j)$ the data observed in areas $B_j$, then we can combine both types of data as follows
\begin{equation}\label{smm}
\begin{split}
       & Y(\vect{s}_i) = \mu(\vect{s}_i) + \phi(\vect{s}_i) + \epsilon(\vect{s}_i), \\
       & Y(B_j) = \frac{1}{B_j} \int_{\vect{u} \in B_j}  \left( \mu(\vect{u}) + \phi(\vect{u}) \right) d\vect{u}+ \epsilon(B_j),
 \end{split} 
\end{equation}
where $\mu(\cdot)$ is the large-scale component of the model that can be explained using covariates, and $\epsilon(\cdot)$ is independent Gaussian noise.

\subsection{Model-based geostatistics under preferential sampling}

Model \eqref{smm}, however, does not take into account the potential preferential sampling that could occur if the sampling locations of the point observations 
are concentrated in regions where the studied phenomenon is assumed 
to be larger (or smaller) than average.
To overcome this issue,
\citet{diggle2010geostatistical} proposed a model that included a latent process $\phi(\cdot)$ shared both by the sampling locations and the measured values.
In this model, the sampling locations of the point observations are modeled using a log-Gaussian Cox process (LGCP) \citep{cox1955some, moller1998log}.
Specifically, the sampling locations 
are modeled by a Poisson process with a varying intensity, which is itself a stochastic process of the form 
\begin{equation*}
\log(\lambda(\vect{\vect{s}})) = 
    \alpha(\vect{s}) + \gamma \phi(\vect{\vect{s}}),
\end{equation*}
where $\alpha(\vect{s})$ is the large-scale component that could include covariates, and $\phi(\vect{s})$ is a zero-mean Gaussian random field (GRF).
Then, the observations are modeled as
\begin{equation} \label{two_com}
    Y(\vect{s}) = \mu(\vect{s}) + \phi(\vect{\vect{s}}) + \epsilon(\vect{s}),
\end{equation}
where $\mu(\vect{s})$ is the large-scale component, and $\epsilon(\vect s)$ is independent Gaussian noise. In this setting, the parameter $\gamma$ controls the degree of preferentiality.

\subsection{Integration of spatially misaligned data under preferential sampling}

It is important to emphasize that while preferential sampling influences the selection of sampling locations, it does not alter the true response surface $Z(\vect s)$. In other words, the selection of the response model is independent of the choice of sampling locations \citep{gelfand2012effect}.  What we can infer about the underlying response surface is indeed influenced by the specific locations chosen for sampling. This distinction is vital in understanding the impact of sampling on data analysis.  

Here, we propose a Bayesian hierarchical model to integrate spatially misaligned data adjusting by preferential sampling by combining the Bayesian melding model, and the shared model for preferential sampling. 
Let the true surface $Z(\vect s) = \mu(\vect{s}) +  \phi(\vect{s}) $  be observed up to white noise at point-level, where $\mu(\cdot)$ is the fixed effect, and $\phi(\vect s)$ is an isotropic GRF.   

The model, which we call ``PSmelding'' in the following sections, is specified as follows:

\begin{itemize}

\item \textbf{Response model:}

Given a point process $\mathcal{S}$ that models the sampling locations $\vect s = \{\vect{s}_1, \dots, \vect{s}_I\}$, we denote by $Y(\vect{s}_i)$ the observed point data. Moreover, we can define $Y(B_j)$ as the areal data at $B_j$, where $\vect \mathcal \vect B = \{ B_1, \dots, B_J\}$ represents the set of analyzed subregions. Under this setting, we assume the following hierarchical model

\begin{equation}\label{bmeld}
\begin{split}
& Y(\vect{s}_i)  \ | \ \mu(\cdot), \phi(\cdot), \mathcal{S}, \tau_s^2 
\sim N( \mu(\vect{s}_i) +  \phi(\vect{s}_i), \tau_s^2), \\
& Y(B_j)  \ | \ \mu(\cdot), \phi(\cdot), \tau_B^2 
\sim N\left( \frac{1}{B_j}\int_{\vect u \in B_j} (\mu(\vect u) + \phi(\vect u)) d\vect u, \tau_B^2 \right).
\end{split}
\end{equation}
From Equation \eqref{bmeld}, notice that the corresponding integral is defined over the subregion $B_j$.


    \item \textbf{Preferential sampling model:}

The set of observed locations $\vect s = \{\vect{s}_1, \dots, \vect{s}_I\}$ is assumed to be a realization of an inhomogenous Poisson point process  (IPPP) $\mathcal{S}$, with intensity function $\lambda(\mathbf{s})$. 

Motivated by \cite{diggle2010geostatistical},
we assume that $\mathcal{S}$ is a log-Cox Gaussian process, such that the logarithm of the intensity function can be modeled with informative covariates $\vect X(\vect s)$ (e.g., population density) and coefficients $\vect \alpha$. That is,
\begin{equation}\label{ippp}
\begin{split}
\mathcal{S} \sim \text{Poisson Process}(\lambda(\vect{s})) \\
\log(\lambda(\vect{s})) = \vect{\alpha} \vect X(\vect s) + \gamma \phi(\vect s)
\end{split}
\end{equation}
If the informative covariates can fully explain the spatial dependency between $\mathcal{S}$ and the true spatial random field, then $\gamma = 0$ and Model \eqref{ippp} is equivalent to the model proposed by \cite{gelfand2012effect}. 

\end{itemize}

Throughout this paper, we will assume that the underlying Gaussian random field is isotropic and stationary with a Mat\'{e}rn covariance structure. The Mat\'{e}rn covariance function is defined as
\begin{equation} \label{matern}
    C(||h||)  = \frac{\sigma^2}{2^{\nu-1}\Gamma(\nu)}
		\left(\kappa \norm{h} \right)^\nu K_{\nu}
		\left( \kappa \norm{h}\right),
\end{equation}
where $||h||$ denotes the distance between locations,
$\sigma^2$ is the variance when $||h||=0$, and
$\nu$ is the smoothing parameter.
The scale parameter $\kappa$ controls the spatial dependence. The Mat\'{e}rn covariance function can also be expressed as a function of the spatial range $\rho = \sqrt{8 \nu}/\kappa$, which corresponds to the distance at with the correlation is close to 0.1 \citep{lindgren2011}. 

As shown in \citet{zhang2004inconsistent}, the parameters of a Gaussian process with isotropic Mat\'{e}rn covariance function at dimension lower than four are not all ``fixed-domain-asymptotic.'' Instead, in fixed domains, only the microergodic parameter $\theta = \sigma^2 \kappa^{2\nu}$
(or continuous functions of it) can be consistently estimated. Therefore, in the following sections, we will consider the microergodic parameter $\theta$ to assess the performance of the proposed model.

\section{Bayesian inference with INLA and SPDE} \label{sec:INLASPDE}

\subsection{INLA}

The integrated nested Laplace approximation (INLA) \citep{rue2009approximate} uses analytical approximations and numerical integration algorithms for Bayesian inference in latent Gaussian models. INLA accelerates statistical inference and estimation of the marginal distributions of parameters, and has become an alternative to Markov Chain Monte Carlo (MCMC) for such models \citep{rue2009approximate,rue2017bayesian}.
INLA offers broad applicability for modeling spatial and spatio-temporal phenomena, with a specific model structure defined as
\begin{equation}
    \begin{split}
        & y_i|\phi, \vect \theta \sim \pi(y_i|\phi, {\vect \theta}),\ \forall i, \\
        & {\phi}|{\vect \theta} \sim N({\mu(\vect \theta)}, {Q(\vect \theta)}^{-1}),\\
        & {\vect \theta} \sim \pi({\vect \theta}),
    \end{split}
\end{equation}
where $i=1, \ldots, I$, and the observed data ${\vect y} = (y_1, \ldots, y_I)$ are commonly modeled by a distribution belonging to the exponential family, with mean $\mu_i = g^{-1}(\eta_i)$. Specifically, $\eta_i$ is the linear predictor that may include the effects of various covariates and random effects. For example, it may consist of an intercept $\alpha$, coefficients ${\beta_k}$ of covariates ${z_{ki}}$, and random effects ${f^{(j)}(\cdot)}$ defined in terms of covariates ${u_{ji}}$.

The latent Gaussian field $\phi$ with mean ${\mu(\vect \theta)}$ and precision matrix ${Q(\vect \theta)}$ is conditioned on hyperparameters ${\vect \theta}$, which follow a prior distribution $\pi({\vect \theta})$. The INLA methodology, with the aid of SPDE \citep{lindgren2011}, can be used to analyze geostatistical observations assumed to be generated from a Gaussian random field with a Mat\'{e}rn covariance function, as presented in the following section.

\subsection{SPDE}

Within the INLA framework, \texttt{R-INLA} \citep{rue2009approximate} also offers support for fitting spatial models that include Gaussian random fields with Mat\'{e}rn dependence structure. \citet{whittle1963stochastic} showed that a Gaussian field with Mat\'{e}rn covariance function can be written as the solution of the following
stochastic partial differential equation (SPDE):
\begin{equation} \label{spde}
(\kappa^2 - \Delta)^{\alpha/2}(\tau \phi(\vect{s})) = \mathcal{W}(\vect{s}),
\end{equation}
where $\Delta = \sum_{i=1}^2 \frac{\partial^2}{\partial x^2_i}$ is the Laplacian, ${\phi}$ is a Gaussian random field, and $\mathcal{W}(\vect{s})$ is a Gaussian spatial white noise process.
$\kappa >0$ represents the scale, $\tau$ controls the variance,
and $\alpha$ is the smoothness of the Gaussian random field.
\citet{lindgren2011} derived, using a finite element method (FEM), a compacted representation with Markov properties for the solution of Equation \eqref{spde}, where ${\phi}$ is approximated by a weighted basis-function expansion, and the joint distribution for the weights is a Gaussian Markov random field (GMRF). INLA-SPDE uses a triangulation of the domain and piecewise linear basis functions in two dimensions.
The representation of the expansion is as follows
\begin{equation} \label{ffe}
   \phi(\vect{s})=\sum_{k=1}^m \omega_k(\vect{s}) x_k, 
\end{equation}
where $m$ is the number of vertices of the triangulation,
$\omega_k(\cdot)$ represents compactly-supported piecewise linear functions defined on each triangle equal to 1 at vertex $k$, and equal to 0 at the other vertices, and $\{x_k\}_{k =1}^{m}$ are zero-mean Gaussian distributed weights.
The precision matrix of the weights is sparse due to Markov properties.

\subsection{Log-Gaussian Cox process fitting using INLA and SPDE}

A common model to describe an observed point pattern as in our sampling scheme is referred to as log-Gaussian Cox process (LGCP).
This process has a varying intensity that depends on a Gaussian random field $\phi$, such the log-intensity is defined as $\log(\lambda^{\star}(\vect s)) = \phi(\vect s)$.
 \citet{illian2012toolbox} developed a fast and flexible framework for fitting LGCP using INLA that approximates the true LGCP likelihood on a regular lattice over the observation window. Later, \citet{simpson2016going}  extended the numerical inference for LGCP processes to the non-lattice case, leading to greater flexibility and computational efficiency.  
In particular, the log-likelihood of a LGCP is defined as
\begin{equation} \label{ll_pp}
    \log\pi(\vect y\mid  \phi) = \abs{\mathcal{D}} - \int_{\mathcal{D}} \exp(\phi(\vect{s}))  d \vect{s} + \sum_{i=1}^n \phi(\vect{s}_i).
\end{equation}

While the SPDE model can compute the sum term exactly, the stochastic integral $\int_{\mathcal{D}} \exp({{\phi}(\vect{s})})d\vect{s}$ must be approximated by a sum.
\citet{simpson2016going} approximated the integral in Equation \eqref{ll_pp} by the deterministic integration rule of the form $\int_\Omega f(\vect s) d \vect s \approx \sum_{i=1}^p \tilde{\alpha}_i f(\tilde{\vect s}_i)$ in the following manner
\begin{equation} \label{approx_likelihood}
\begin{split}
 \log\left[\pi(\vect y\mid { \phi})\right] &\approx C - \sum_{i=1}^p \tilde{\alpha}_i\exp\left[\sum_{k=1}^m x_k \omega_k(\tilde{\vect s}_i)\right]+ \sum_{i=1}^n \sum_{k=1}^m x_k\omega_k(\vect s_i)\\
&= C - \tilde{\vect{\alpha}}^{\top} \exp(\vect{A_1} \vect{x}) + \vect 1^{\top} \vect{A_2} {\vect x},
\end{split}
\end{equation}
where $C$ is a constant, $[A_1]_{ik} = \omega_k(\vect{\tilde{s}}_i)$ is a projector matrix containing the values of the latent Gaussian model at the integration nodes $\vect{\tilde{s}}_i$, and $[A_2]_{ik} = \omega_k(\vect{s}_i)$ evaluates the latent Gaussian field at the observed points $\vect{s}_i$. 

\begin{figure}[!ht]
    \centering
    \includegraphics[width = 0.6\textwidth]{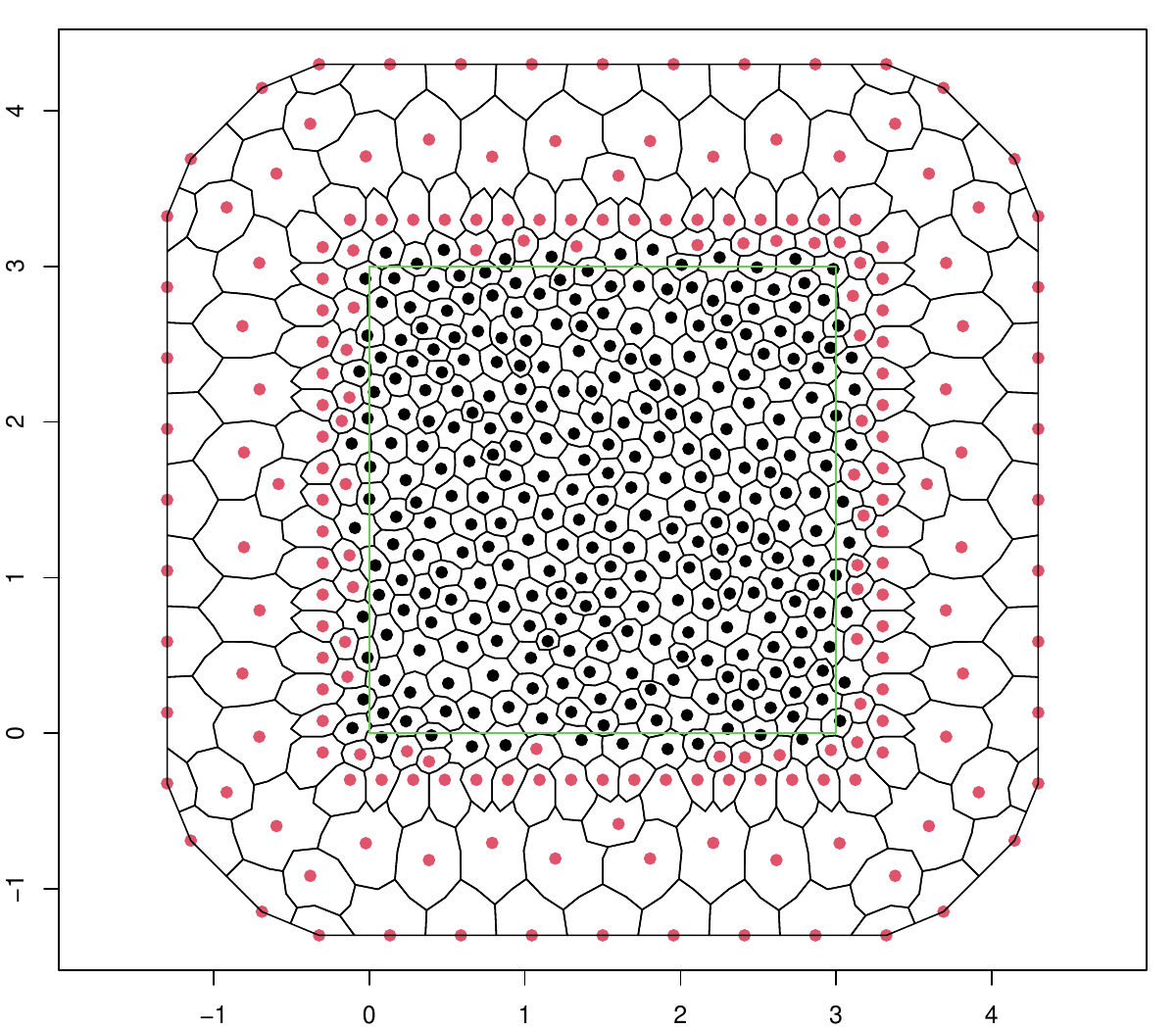}
    \caption{Dual mesh constructed by joining the centroids of the primal mesh. Black nodes represent the centroids and
    red nodes the integration points with zero weight.
    The green square is the area of interest.}
    \label{fig:dual_mesh}
\end{figure}

Using numerical integration schemes proposed by \citet{baddeley2000practical}, \citet{simpson2016going} construct pseudo-observations, $\vect y = (\vect 0_{p \times 1}^{\top} ,\vect 1_{N \times 1}^{\top})^{\top}$, and approximate the likelihood using the form $\pi(\vect y \mid {\phi}) \approx \vect C\prod_{i=1}^{n + p} \eta_i^{y_i} e^{-\alpha_i u_i}$, where $\log \vect \eta = (\vect{x}^{\top} \vect{A_1}^{\top}, \vect{x}^{\top} \vect{A_2}^{\top})^{\top}$, $\alpha_i$ and $u_i$ correspond to the intensity function and the basis function evaluated at the $i$-th point, respectively. This likelihood approximation is similar to that of observing $N+p$ conditionally independent Poisson random variables with means $\alpha_i u_i$ and observed values $y_i$. Aiming at fully specifying the model, it is necessary to define an integration scheme that can be employed in Equation \eqref{approx_likelihood}. One of the simplest approaches is to assign a region $V_i$ to each node within the mesh, where the basis function $\phi_i(\vect s)$ takes on a greater value than any other basis function, as in Figure \ref{fig:dual_mesh}. The associated integration rule involves setting $\tilde{\vect s}_i$ as the location of the node, and $\tilde{\alpha}_i = |V_i|$ as the volume of the dual cell.
Our work uses this algorithm to estimate the model parameters considering preferential sampling.

\subsection{Spatially misaligned data modeling using INLA and SPDE}

As shown in \citet{Moraga2017},
spatial data available at different resolutions can be combined by using a Bayesian melding model that can be fitted using the INLA-SPDE framework.
In this approach, point data $Y(\vect{s}_i)$ are realizations of the shared latent Gaussian random field at sites $\vect{s}_i$,
and areal data $Y(B_j)$ can be represented as stochastic integrals within the areas.
To reduce the computational cost, \citet{Moraga2017} used the fact that the weighted sum of the vertices in the triangulated mesh is the approximation of the integral in the area.
Thus, if $\vect{A}$ represents the projector matrix that maps the Gaussian Markov random field from the observations to the triangulation nodes, then the areal observations can be expressed as
\begin{equation}
       E(Y(B_j)) =  \int_{\vect{s} \in B_j} \phi(\vect{s}) d\vect{s}
         \approx \sum_{i = 1} ^n \sum_{k=1}^m x_k \omega_k(\tilde{\vect{s}}_i) = \vect 1^{\top} \vect{A} {\vect x}.
\end{equation}

\begin{figure}[!ht]
    \centering
    \includegraphics[width = 0.6\textwidth]{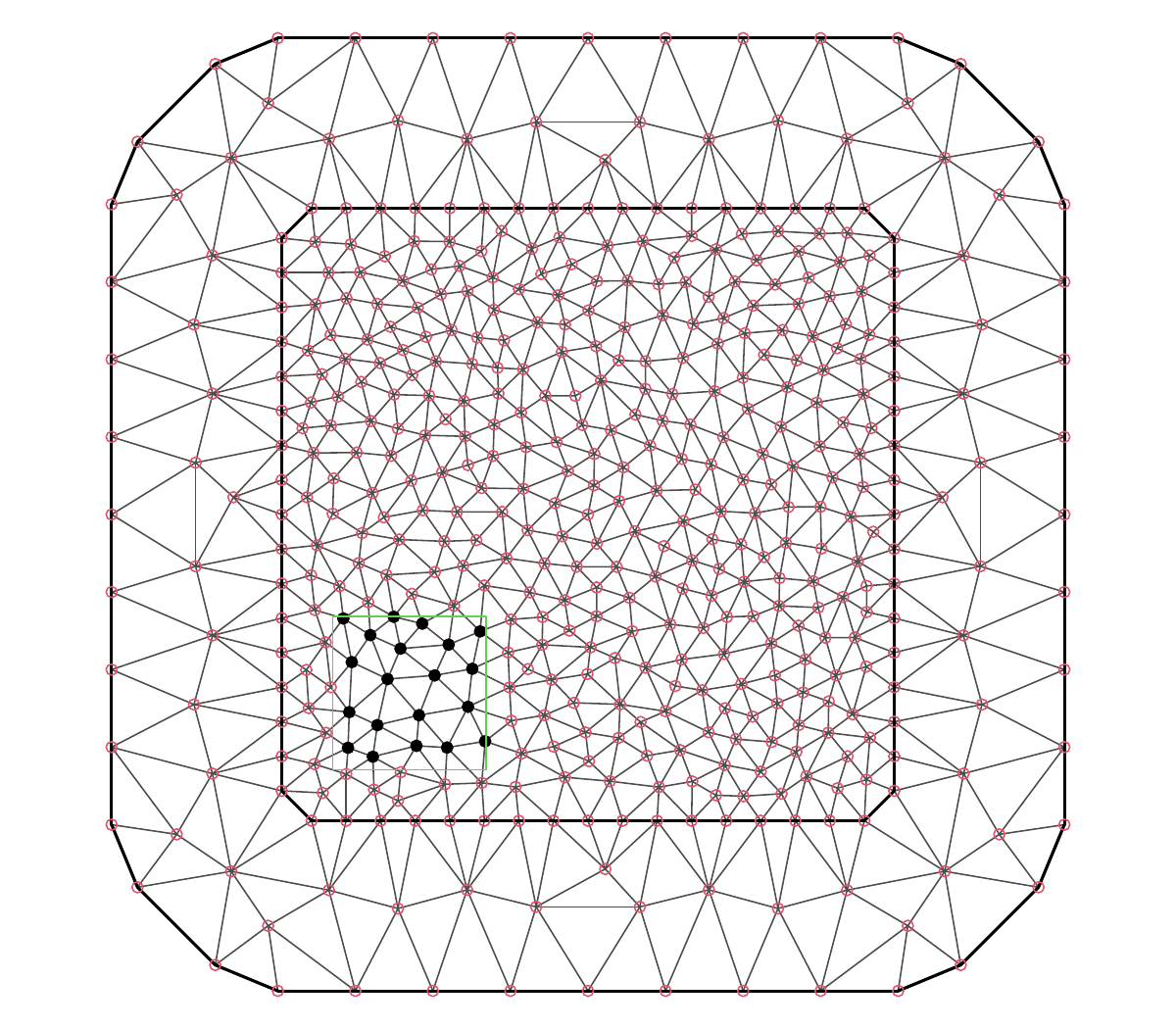}
    \caption{Triangulated mesh used in the INLA-SPDE approach. The green square represents one of the areal observations.
    In the row of the projector matrix corresponding to this area, the $n$ black vertices within the area are weighted $1/n$. The vertices outside the area are given zero weight.}
    \label{fig:mesh}
\end{figure}

Figure \ref{fig:mesh} illustrates the triangulation vertices used to compute the projector matrix when combining point and areal observations.
For a given area $B_j$, the corresponding $j$-th row of the projector matrix has elements that satisfy
$\sum_{i = 1} ^m [A]_{ij} = 1$. 
For vertices outside the area,
$[A]_{ij} = 0$.
For vertices inside the area,
$[A]_{ij} = 1/m_{B_j}$, where $m_{B_j}$ is the number of points contained in $B_j$. 
Note that, while \citet{Moraga2017} uses equal weights for areal observations, it is also possible to use different values in situations where other factors, e.g., population density, need to be taken into account.

In our implementation, we assume that preferential sampling only happens in geostatistical data. Thus, the projector matrix of observations $\vect{A_2}$ in Equation \eqref{approx_likelihood} maps point observations only. In the response model, the projector matrix projects both point and areal level data. 

\section{Simulation study}
\label{s:sim}

We conduct a simulation study to analyze the performance of our proposed melding approach adjusting for preferential sampling (PSmelding) in comparison with the melding approach without preferential sampling adjustment (Melding) and a geostatistical model that only uses point data adjusting for preferential sampling (PSgeo).

Specifically, we assess the
bias that is produced by ignoring preferential sampling when using a melding approach to combine spatially misaligned data, and how spatial data fusion model reduces the risk of misspecification when preferential sampling does not exist.

\subsection{Data simulation}

In all simulation scenarios, we assume that the region of interest is a unit square.
We generate random surfaces from Gaussian random field with Mat\'{e}rn covariance functions.
We then obtain point observations of the simulated surfaces at 100 and 250 randomly generated locations. We also obtain areal observations by averaging the true surfaces at cells of a regular grid consisting of 4, 25, or 100 subregions. \ruiman{We design two sampling procedures.} In particular, the \ruiman{first} sampling process for point-level data is from Model \eqref{ippp}, \ruiman{whereas the second sampling strategy selects observations larger than a fixed threshold.
The point-level observations are sampled} with the intercept as the only fixed effect, and the responses are simulated from Model \eqref{bmeld}. That is, our simulation assumes the areal and point responses are from a common spatial random field and informative covariates are not available to capture the dependency between sampling process and the spatial random field. This is because we focus on how preferential sampling affects inference and how our proposed model reduces its effect. Also, we acknowledge that assuming the responses for both processes come from a common model may be unreasonable in some real-world settings---thus, in Section \ref{s:applicaton}, we relax this assumption.

Then, we fit the Melding, PSmelding and PSgeo models to different configurations of the observed data, and obtain predictions at locations of a 50 $\times$ 50 regular grid. We assess the performance of each of the models by comparing the true and simulated surfaces using several criteria.
For each of the scenarios, we generate 100 spatial surfaces to have stable results.

We simulate data according to six scenarios. In particular, we set $\mu(\vect s) = 0$. Also, regarding the intensity function of the LGCP, we set $\alpha(\vect s) = 0.05$. The variance of the random noise (nugget effect) of the point and areal observations is set to 0.1, indicating precision $\tau_s$ = $\tau_B = 10$.
The shared spatial random effect $\phi(\cdot)$ is assumed to be a zero-mean Gaussian random field with isotropic Mat\'{e}rn covariance function parameterized as in Equation \eqref{matern}.
For all scenarios, we set the parameters of the covariance function $\nu$ = 1 and $\sigma$ = 1.
We set $\kappa$ equal to $\sqrt{2 \nu}/\texttt{scale}$, with  $\texttt{scale} = 0.05, 0.1, 0.2$.
This implies values of the microergodic parameter
equal to
$\theta = \sigma^2 \kappa^{2\nu} =  2\sigma^2/\texttt{scale} ^{2}$ equal to 800, 200 and 50, as seen in Table \ref{tab:1}.
A more interpretative parameter is the spatial range which is equal to $\rho = \sqrt{8 \nu}/\kappa = \sqrt{8 \nu} \times \texttt{scale}/\sqrt{2 \nu} =  2 \times \texttt{scale}$, which corresponds to the distance at with the correlation is near 0.1.
Finally, the preferential sampling degree parameter $\gamma$ is set to 0 to simulate non-preferential sampling scenarios, or to 1 to represent preferential sampling scenarios.

\begin{table}[ht] 
    \centering
    \caption{Parameter values used in each of the simulation scenarios. There are six scenarios arising as combinations of
    microergodic parameter ($\theta$) and preferential degree ($\gamma$).}
     \begin{tabular}{l c c c c c c   }
     Scenario  &  1  & 2 & 3 & 4 & 5 & 6 \\
    \hline
    $\rho$ & 0.1 & 0.2 & 0.4 & 0.1 & 0.2 & 0.4   \\
    $\theta$ & 800 & 200 & 50  & 800 & 200 & 50 \\
    $\gamma$ & 0 & 0 & 0 & 1 & 1 & 1  
    \end{tabular}
    \label{tab:1}
\end{table}

\subsection{Model assessment}

Aiming at assessing the results, we report the mean squared error (MSE), the mean absolute error (MAE) and the Wasserstein distance (WD) \citep{kantorovich1960mathematical} obtained for each of the simulated scenarios.
The MSE is calculated by averaging the squared differences between the predictive posterior mean and the true spatial surface, whereas the MAE averages the absolute value of the differences between the predictive posterior median and the true surface.
Since both MSE and MAE ignore the variability of the predictive posterior distribution, we also use the WD as a criterion to account for the whole predictive posterior distribution.

The WD
has been used
to assess the robustness and statistical properties of distributions arising
in different fields 
including image analysis, metereology and astrophysics
\citep{engquist2013application,zhao2018data, mohajerin2018data,sun2022optimizing}.

\citet{givens1984class} provide the closed form of WD to compare Normal distributions. Specifically, if $ \pi _{1}={\mathcal{N}}(\vect \mu_{1},\vect C_{1})$ and 
 $\pi _{2}={\mathcal{N}}(\vect \mu_{2},\vect C_{2})$ are two non-degenerate Gaussian measures on 
$\mathbb {R} ^{n}$ with respective expected values 
$\vect \mu_{1},\vect \mu_{2}\in \mathbb {R} ^{n} $  and covariances $\vect C_{1}, \vect C_{2} \in \mathbb {R} ^{n\times n}$, the WD can be expressed as
\begin{equation*} \label{wassertein}
    W_{2}(\pi_{1},\pi_{2})^{2}=\| \vect \mu_{1}- \vect \mu_{2}\|_{2}^{2}+\mathop {\mathrm {trace} } {\bigl (}\vect C_{1}+\vect C_{2}-2{\bigl (}\vect C_{2}^{1/2}\vect C_{1}\vect C_{2}^{1/2}{\bigr )}^{1/2}{\bigr)}.
\end{equation*}

In our case, we measure the conditional predictive posterior distribution $\hat{\pi(y_i|\vect{y}})$ and the true distribution of the $\pi(y_i|\vect{y})$ point-wise and calculate the sum of the WD for all prediction sites $\vect{y} = (y_1, \ldots, y_n)$. In this setting, the WD can be written as
\begin{equation*}
    W_{2}(\hat{\pi}(y_i|\vect{\hat{y}}),\pi(y_i|\vect{y}))^{2}=\|\hat{\mu}-y_i\|_{2}^{2}+\mathop {\mathrm {trace} } {\bigl (}\hat{\sigma^2} +\sigma^2 -2\hat{\sigma}\sigma{\bigr)},
\end{equation*}
where $\hat{\mu_i}$ and $y_i$ are the estimated and true mean of predictive posterior for site $i$, $i=1,\ldots,n$, and $\hat{\sigma}$ and $\sigma$ are the estimated and true variances, respectively. 

In addition, we also evaluate the parameter estimates by using the average of the 95\% coverage probabilities (CP), and the average interval score of 95\% prediction intervals \citep{gneiting2007strictly}.
The 95\% CP are defined as the proportion of times that the real parameter is within the 95\% credible interval obtained in each of the simulations.
The interval score is defined as
\begin{equation*}
    \mbox{Inscore}_\alpha\left(I ; y \right) = (I_{u}-I_{l})+\frac{2}{\alpha}\left(I_{l}-y\right)_{+}+\frac{2}{\alpha}\left(y-I_{u}\right)_{+},
\end{equation*}
where
$y$ is the true parameter and
$I = [I_{l},I_{u}]$ is  the $100(1-\alpha)$\% credible interval, with $\alpha = 0.05$ in our study.
$(x)_{+} = x$, if $x \geq 0$, and $(x)_{+} = 0$, if $x < 0$.
A smaller interval score is preferable as the score rewards high coverage and narrow intervals. 

\subsection{Results}

\subsubsection{Surface estimates}

Table \ref{tab:eva_de_100} shows the performance of the three different models, namely, Melding, PSmelding, and PSgeo in terms of the mean and 95\% quantile intervals of MSE, MAE, and WD in scenarios that combine 0, 4, 25, and 100 areas with 100 point observations.
These results correspond to data generated with microgeodic parameter $\theta$ = 200 implying a spatial range $\rho$ = 0.2 both in preferential and non-preferential scenarios. The results for the remaining simulated scenarios are in the Supplementary Material.

\begin{table}[ht]
\caption{Mean of the scores and 95\% quantile intervals obtained with Melding, PSmelding and PSgeo models in scenarios that combine 100 point observations with 0, 4, 25, and 100 areas under preferential sampling and non-preferential sampling scenarios.
Results corresponding to data generated with spatial range $\rho$ = 0.2 and microgeodic parameter $\theta$ = 200. In all scenarios, and for each criterion, if there is a unique best model (based on the reported mean), we highlight it in bold.}
\centering
\resizebox{\columnwidth}{!}{%
\begin{tabular}[t]{l l l ll l | l l l}
\hline
\multicolumn{3}{c}{ } & \multicolumn{3}{c}{Preferential sampling} & \multicolumn{3}{c}{Non-preferential sampling} \\
\cline{4-6} \cline{7-9}
Model & Areas & $\rho$ & MSE  & MAE  & WD  & MSE & MAE& WD \\
\hline
Melding & 4 & 0.2 & 0.47 (0.3; 0.83) & 0.52 (0.42; 0.68) & 0.84 (0.65; 1.11) & \textbf{0.39 (0.3; 0.56)} & \textbf{0.48 (0.43; 0.56)} & 0.8 (0.63; 1.07)\\
Melding & 25 & 0.2 & 0.26 (0.21; 0.33) & 0.39 (0.36; 0.44) & 0.56 (0.44; 0.72) & 0.34 (0.26; 0.46) & 0.46 (0.4; 0.52) & 0.71 (0.57; 0.88)\\
Melding & 100 & 0.2 & 0.14 (0.11; 0.17) & 0.29 (0.26; 0.32) & 0.28 (0.24; 0.34) & 0.22 (0.18; 0.25) & 0.37 (0.33; 0.4) & 0.47 (0.41; 0.55)\\
\hline
PSmelding & 4 & 0.2 & \textbf{0.35 (0.26; 0.47)} & \textbf{0.45 (0.39; 0.53)} & \textbf{0.66 (0.54; 0.78)} & 0.4 (0.3; 0.6) & 0.49 (0.43; 0.59) & 0.8 (0.63; 1.09)\\
PSmelding & 25 & 0.2 & 0.26 (0.21; 0.33) & 0.39 (0.36; 0.44) & \textbf{0.5 (0.4; 0.61)} & 0.34 (0.25; 0.46) & 0.46 (0.4; 0.52) & \textbf{0.7 (0.57; 0.85)} \\
PSmelding & 100 & 0.2 & 0.14 (0.11; 0.17) & 0.29 (0.26; 0.32) & \textbf{0.27 (0.22; 0.31)} & 0.22 (0.18; 0.25) & 0.37 (0.33; 0.4) & 0.47 (0.41; 0.55)\\
\hline
PSgeo & 0 & 0.2 & 0.4 (0.27; 0.6) & 0.48 (0.4; 0.58) & 0.71 (0.56; 0.9) & 0.41 (0.31; 0.61) & 0.49 (0.44; 0.58) & 0.84 (0.65; 1.17)\\
\hline
\end{tabular}%
\label{tab:eva_de_100}
}
\end{table}

For the preferential sampling scenario and for a specific number of areal observations, the PSmelding model presents the best performance, as indicated by the lowest MSE, MAE, and WD values. 
When the number of areas is relatively large (i.e., equal to 100), the MSE and MAE for both the PSmelding and Melding models are the same since the areal data has higher influence than the points on the posterior distributions obtained.
However, the PSmelding model presents lower WD values with narrower quantile intervals.
Moreover, the PSmelding model has lower MSE, MAE, and WD values than the PSgeo model even when the number of areal data is small, and with narrower quantile intervals.

For the non-preferential sampling scenario,  the PSmelding and Melding models have similar MSE, MAE and WD values. 

For both preferential and non-preferential sampling scenarios, we observe that, as the number of areas increase, the difference between the PSmelding and PSgeo models becomes larger, indicating that a model that combines point- and areal-level data has better performance than a model that just uses point-level data.

\ruiman{The results for the misspecified scenarios are shown in Table 3 in Supplementary Material. PSmelding outperforms the other two models when the cut-off value for selecting point-observations is 1.5.} These findings suggest that the PSmelding model is a promising approach for addressing the challenges of spatial prediction in presence of preferential sampling (here, we discuss the results obtained for a specific value of microergodic parameter, however, similar results are obtained for other simulation scenarios, as shown in the Supplementary Materials).

\begin{figure}
    \centering
    \includegraphics[width = 1\textwidth]{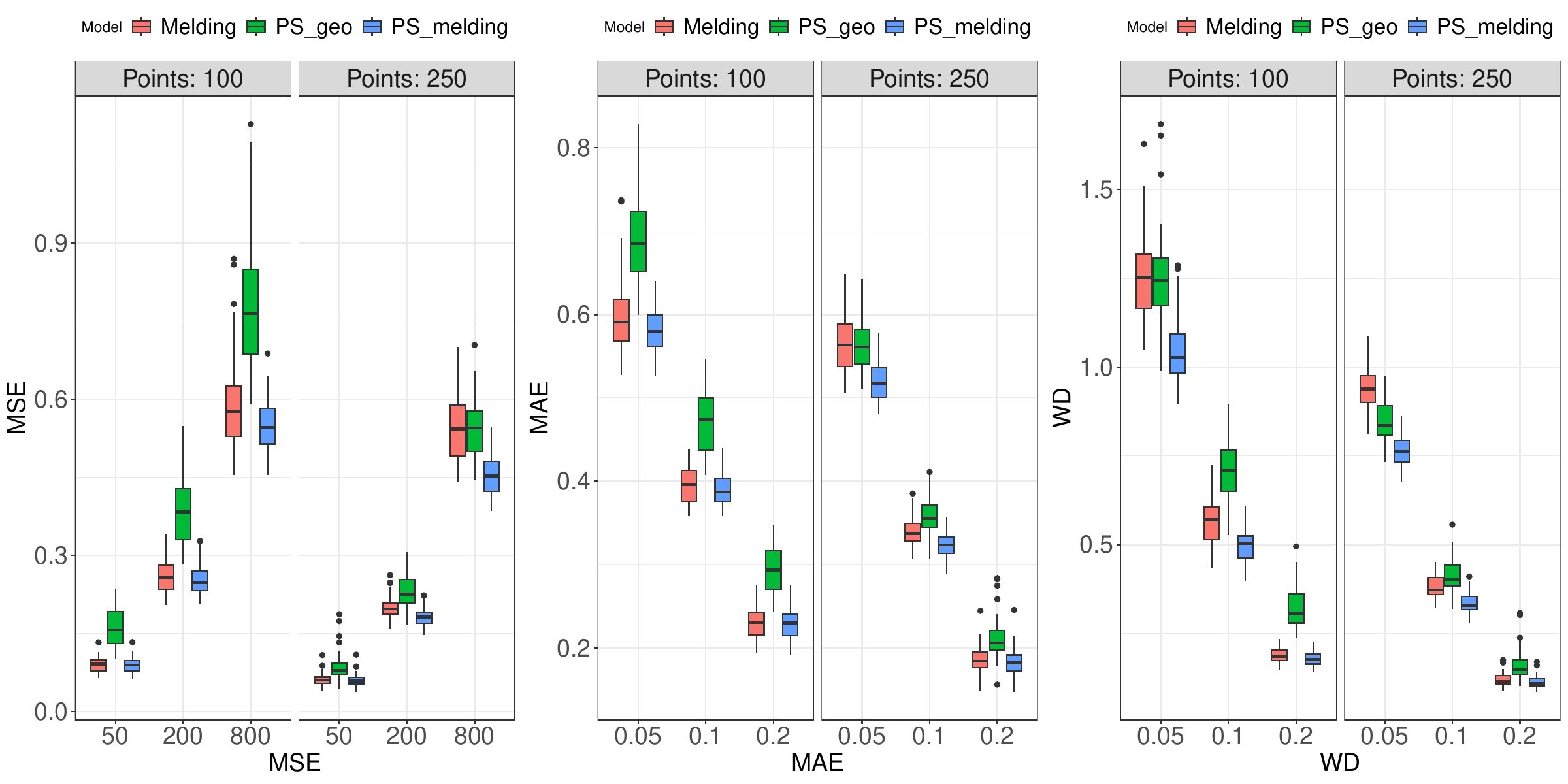}
    \caption{MSE, MAE, and WD values obtained by fitting each of the models when combining 100 or 250 points with 25 areas.
    Data are simulated under preferential sampling using a value of the microergodic parameter $\theta$ indicated in the horizontal axis.}
    \label{fig:pc25_sim}
\end{figure}

Besides investigating the effect of the number of areas in the models performance, we can also assess the results obtained based on different
values for the microerdogic parameter $\theta = \sigma^2 \kappa^{2\nu}$
used to generate the Gaussian random fields.
Figure \ref{fig:pc25_sim} shows the MSE, MAE and WD values obtained when fitting the three models to 100 and 250 points and 25 areal observations obtained in scenarios generated with preferential sampling.
The results for the remaining settings can be found in the Supplementary Material.

Figure \ref{fig:pc25_sim} shows worse prediction performance as the microgeodic parameter increases.
This may indicate that more information is needed
to improve the prediction quality when
the GRF variance increases or when the GRF spatial range decreases.
In addition, the differences between PSmelding and other two models increases as $\theta$ decreases (also, as the number of the points decreases). This indicates that our model improves the prediction ability by merging information from the sampling process, areal data, and point data, especially when preferential sampling occurs with lower spatial dependency and limited geostatistical information. The conclusion is valid for other preferential sampling settings (see Supplementary Material).

\subsubsection{Parameter estimates} 

We also evaluate the estimates of the fixed effect and Gaussian random field parameters.
Tables 6,7 and 8 in Supplementary Materials present the average posterior mean, interval scores (Inscore) and 95\% coverage probabilities (CP) for the fixed effect $\mu$, the microergodic parameter $\theta$, and the preferential degree $\gamma$ obtained when all models are fitted to several combinations of areal and point data.

When estimating the mean $\mu$, the PSmelding model yields a posterior mean closer to the true value ($\mu = 0$), followed by the PSgeo and Melding model. Fusing reliable areal data results in lower Inscore across scenarios, and increasing the spatial range improves posterior mean estimation of $\mu$. PSmelding consistently performs better on Inscore and CP, with the difference between PSmelding and other models decreasing as spatial range increases. The average posterior mode and credible interval scores are indistinguishable at $\rho = 0.4$.

For the microergodic parameter $\theta$, at $\rho = 0.4$ and $\theta = 50$, the three models show similar performance in posterior mean and credible interval. Sampling size has little influence. At $\rho = 0.2$ and $\theta = 200$, increasing the number of points decreases the posterior mean. Also, ignoring preferential sampling (Melding) results in a wider credible interval. The average posterior mode and CP of PSgeo decrease with more points; on the other hand, Inscore remains stable under the same setting. PSmelding achieves a trade-off between credible interval width and coverage.

At $\rho = 0.1$ and $\theta = 800$, all models perform worse due to low spatial dependency. PSgeo has low average Inscore and low CP, Melding has high average Inscore and high CP, and PSmelding outperforms, on average, the other approaches (all based on the posterior mode).

Regarding preferential sampling, PSmelding and PSgeo yield similar posterior means for $\gamma$. Lastly, when $\rho = 0.1$, PSmelding recovers $\gamma$ better than PSgeo based on the CP and Inscore values.

\section{Spatial interpolation of $\text{PM}_{2.5}$ levels in USA }\label{s:applicaton}

Particulate matter 2.5 ($\text{PM}_{2.5}$) is a type of air pollutant consisting of tiny particles with a diameter of 2.5 microns per cubic meter or less that are suspended in the air. These particles can originate from various sources, including vehicle exhausts, industrial emissions, and natural events such as wildfires. Due to their small size, $\text{PM}_{2.5}$ particles can easily penetrate deep into the lungs and enter the bloodstream leading to adverse health effects 
including respiratory and cardiovascular diseases and increased mortality rates \citep{kampa2008human, di2017air, thurston2017joint}.

In the United States of America (USA), the Environmental Protection Agency (EPA) has established National Ambient Air Quality Standards (NAAQS) for $\text{PM}_{2.5}$ to protect public health and the environment. However, despite efforts to control and reduce $\text{PM}_{2.5}$ levels, many areas in the country continue to experience high levels of this harmful pollutant, particularly urban areas and regions with significant industrial activity. 

\begin{figure}[!ht]
\centering
\caption{$\text{PM}_{2.5}$ levels in the USA in 2019 obtained at locations of EPA monitoring sites (left) and grid cells generated by NASA satellites (right).}
\includegraphics[width = 1\textwidth]{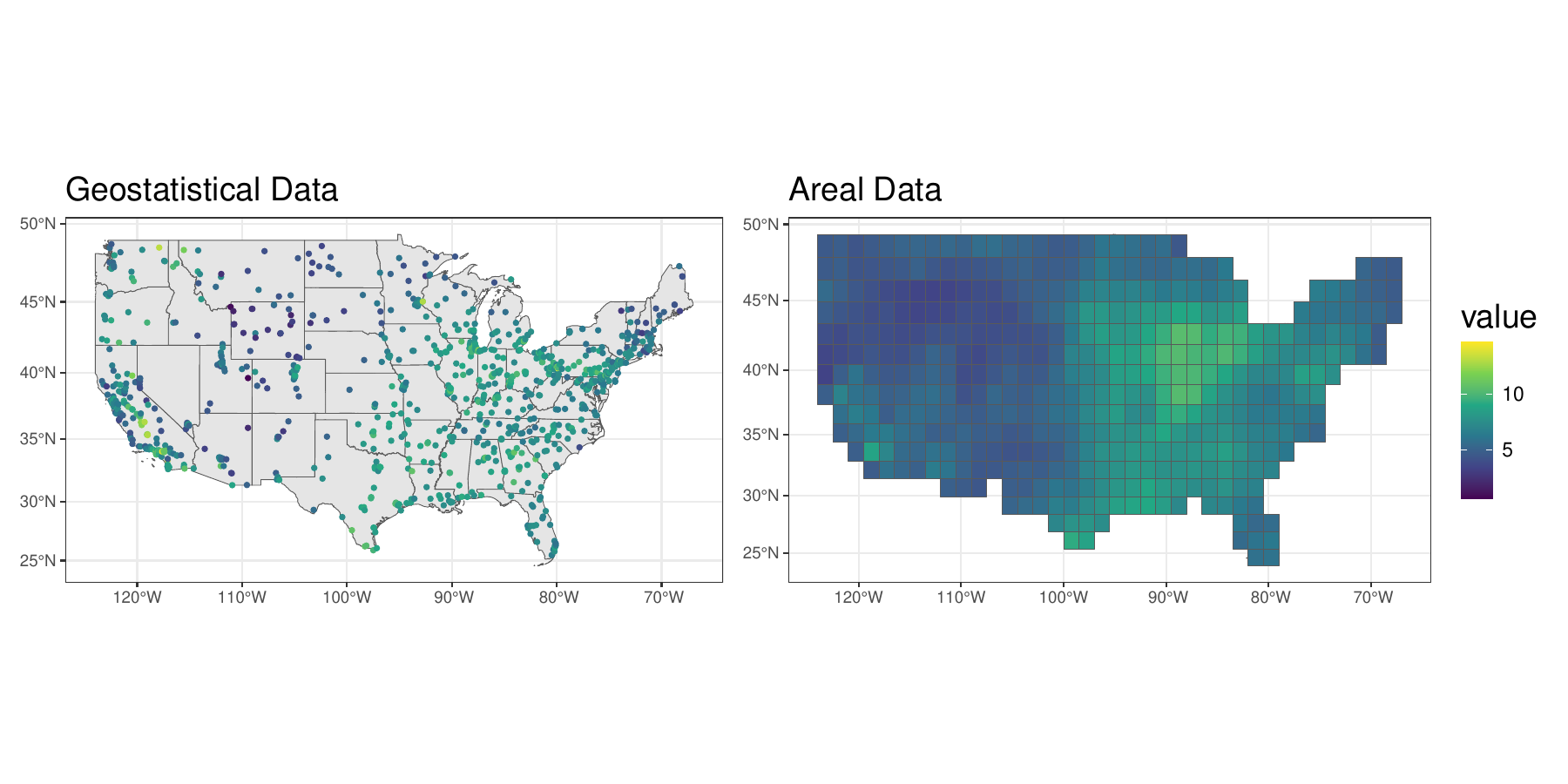}
\label{fig:orgdata}
\end{figure}

Figure \ref{fig:orgdata} shows the more than 900 point data obtained from EPA \citep{epa:2019}, 
and the 280 aggregated areal data at a resolution 25 $\times$ 25 degrees from  \cite{hammer2020global,2019data}. \cite{hammer2020global} estimated annual $\text{PM}_{2.5}$ levels using data from multiple satellites' Aerosol Optical Depth (AOD) measurements, AOD products, ground-based observations, and simulation models. To derive $\text{PM}_{2.5}$ surface values, they firstly applied the intercalibration of the real satellite AOD observations and simulated AOD sources. Then they calculated the estimated $\text{PM}_{2.5}$ from the calibrated AOD sources. Thus, the areal data here is not an aggregation of point data, but we can still observe a strong correlation between two data sources. This motivates us to fit our spatial misaligned model. From the two data sources, we observe that more monitoring sites are located in the eastern part of the US and California, with higher $\text{PM}_{2.5}$ than that of remaining regions.
Thus we can see the evidence for a response-biased sampling design (PS).

In this section, we fit the proposed spatial misaligned model to point and areal level measurements adjusting for preferential sampling on $\text{PM}_{2.5}$ in the USA in 2019. The selected year aims at avoiding 
the influence of COVID-19.

\begin{figure}
    \centering
    \includegraphics[width = 0.6 \textwidth]{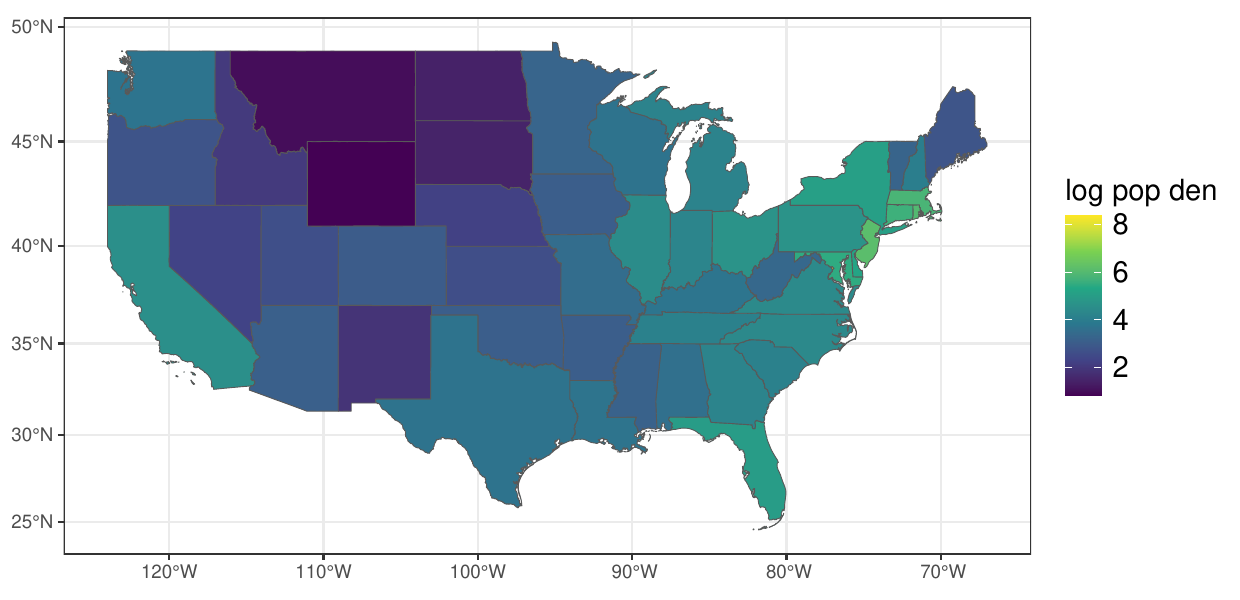}
    \caption{Population density in USA (log-scale).}
    \label{fig:usa_pd}
\end{figure}

To ensure the model captures the real impact of the sampling scheme, as discussed in \cite{diggle2010geostatistical, gelfand2012effect, watson2019general}, we consider an informative covariate $\vect X(\vect s)$ representing the state-level population density from the US Census Bureau \citep{uscensuspopulationdensity}. The population density (log-scale) is shown in Figure \ref{fig:usa_pd}.

In our study, we aim to apply the PSmelding approach to fuse the areal response and potentially preferential-sampled geostatistical observations. As the areal data are no longer observations, we will release the assumption that the two data source are from the same spatial random field. Thus, we implement a more flexible model that allows the areal data and point data to be generated from two different spatial random fields.

In particular, the response model is as follows
 \begin{equation}\label{two_meld}
     \begin{split}
     & Y(\vect{s}_i)  \ | \ \mu(\cdot), \phi(\cdot), \mathcal{S}, \tau_s^2 
      \sim N( \mu(\vect{s}_i) +  \phi(\vect{s}_i), \tau_s^2) \\
        & Y(B_j)  \ | \ \mu(\cdot), \phi(\cdot),\beta, \tau_B^2 
      \sim N\left( \frac{1}{B_j}\int_{\vect u \in B_j} (\mu(\vect u) + \zeta \phi(\vect u)) ) d\vect u, \tau_B^2 \right),
     \end{split}
 \end{equation}
where $\phi(\vect s)$ are GRF with isotropic Matern covariance function, $\mu(\vect s) = \beta_0 + \beta_1 \texttt{pop\_dens}(\vect s)$ are linear combination of intercept and population density times coefficients, and $\zeta$ describes the differences of random effects of two responses. The preferential sampling model is determined as in Equation \eqref{ippp}. 

\ruiman{In Section \ref{s:sim}, we assessed the effectiveness of all models assuming the structure in Equation \eqref{bmeld}, where we showed that accounting for the dependence between the latent field and the sampling process may enhance interpolation accuracy in scenarios with limited areal data. Similarly, we would like to assess Model \eqref{two_meld}.
To do so, we conducted an additional simulation study with parameters identical to those of the previous experiment and set
$\zeta = 0.8$. The results are shown in Table 4 and 5 in the Supplementary Material. From those tables, we can draw similar conclusions as before, that is, the
PSmelding outperforms the other two models when the amount of areal data is relatively small.}

\begin{figure}[!ht]
\centering
\caption{
Rank correlations between the latent field and the $k$-nearest neighbours distances.}
\includegraphics[width = 0.6\textwidth]{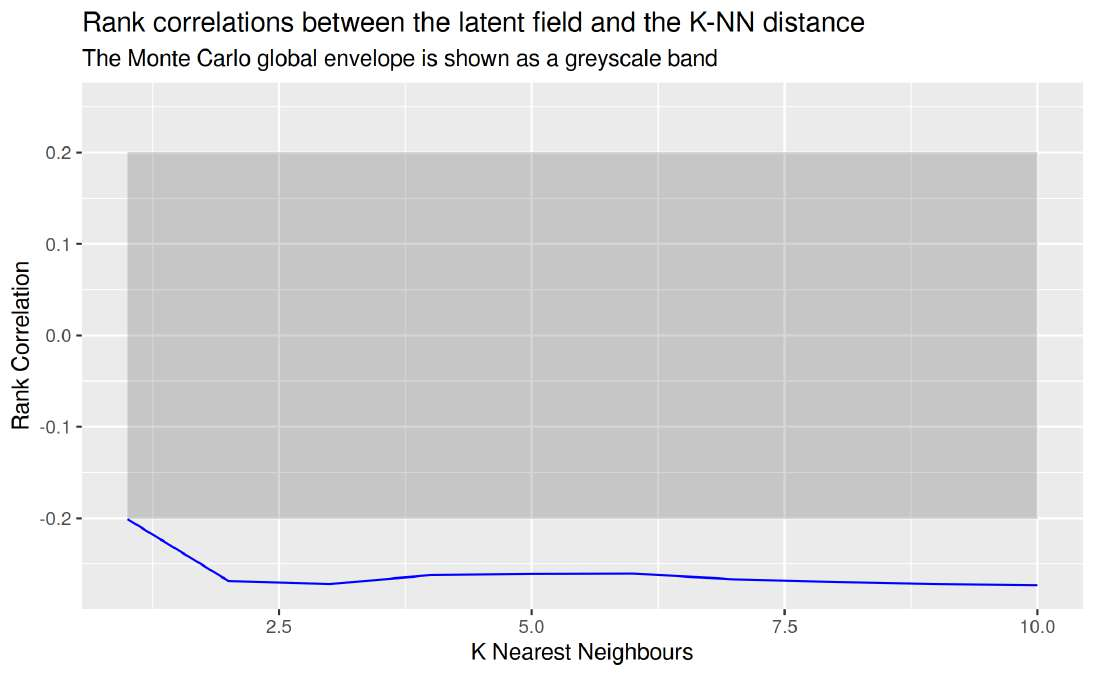}
\label{fig:test}
\end{figure}

\ruiman{A perceptron-based test, proposed by \cite{watson2021perceptron}, is applied to test for preferential sampling, given the population density. The test uses a Monte Carlo algorithm for testing the null hypothesis that there is no correlation between the latent effect and the intensity of the point pattern. Under the null hypothesis, the observation and sampling processes are conditionally independent. Conversely, a positive association between the localized amount of spatial clustering and estimated 
latent field $\phi(\vect s)$ would be expected. The mean of the $k$-nearest neighbours (K-NN) can be used to  capture clustering. 
The test can be modified for covariates, which allow us to test whether the population density controls the preferential sampling.
Figure \ref{fig:test} shows that the observed rank correlation lies outside the Monte Carlo rank envelope for all values of K-NN tested. Thus, we reject the null hypothesis at the $5\%$ level.}

To fit the models, we first project the spatial data which is given in a geographic coordinate reference system into a two-dimensional surface using the Mercator projection with units in kilometers. 
Then, we create a common mesh to fit all the models using the INLA-SPDE approach.
The dual mesh for estimating the intensity surface of the point process is created using the center of the triangles of the original mesh with the same number of the nodes. 
In order to obtain reasonable priors for the parameters of the Mat\'{e}rn covariance function of the Gaussian random field, we first fit the model to areal data only.
Then, we use the posterior means of the parameters to set priors $\mathbb{P}(\sigma > 0.1) = 0.6$, and $\mathbb{P}(\rho < 880 ~\text{km}) = 0.90$, and $\nu = 1$.

Table \ref{tab:psmeld_para_USA} shows the estimated parameters and hyperparameters for the fitted model. \ruiman{The estimated degree of preferentiality $\gamma$ indicates that population density at state-level cannot fully explain the dependency between sampling process.}

\begin{table}[ht]
\caption{Posterior means 95\% credible intervals of the parameters and hyperparameters obtained with PSmelding. $ \beta_0 \text{ and } \beta_1$ are the coefficients in the responses model and $ \alpha_0 \text{ and } \alpha_1$ are the coefficients in the sampling model.}
\centering
\begin{tabular}{rrrrr}
  \hline
 & mean & std & 2.5 \% & 97.5 \% \\ 
  \hline
  $\beta_0$ & 5.68 & 0.33 & 5.05 & 6.34 \\ 
  $\beta_1$ & 0.12 & 0.09 & 0.01 & 0.27 \\
  $\alpha_0$ & -11.83 & 0.18 & -12.19 & -11.47 \\ 
  $\alpha_1$ & 0.57 & 0.05 & 0.48 & 0.66 \\  
  Range for $\phi(\vect s)$ & 495.32 & 57.23 & 396.08 & 621.08 \\ 
   Stdev for $\phi(\vect s)$ & 1.45 & 0.09 & 1.28 & 1.64 \\ 
  $\gamma$ & 0.38 & 0.03 & 0.31 & 0.44 \\ 
  $\zeta$ & 0.85 & 0.05 & 0.74 & 0.95 \\ 
  $\tau_s$  & 0.74 & 0.04 & 0.67 & 0.81 \\
  $\tau_B$ & 0.72 & 0.05 & 0.63 & 0.80  \\
   \hline
\end{tabular}
\label{tab:psmeld_para_USA}
\end{table}

\begin{figure}[!ht]
\centering
\caption{$\text{PM}_{2.5}$ levels in the USA in 2019 interpolated using the PSmelding model.}
\includegraphics[width = 1\textwidth]{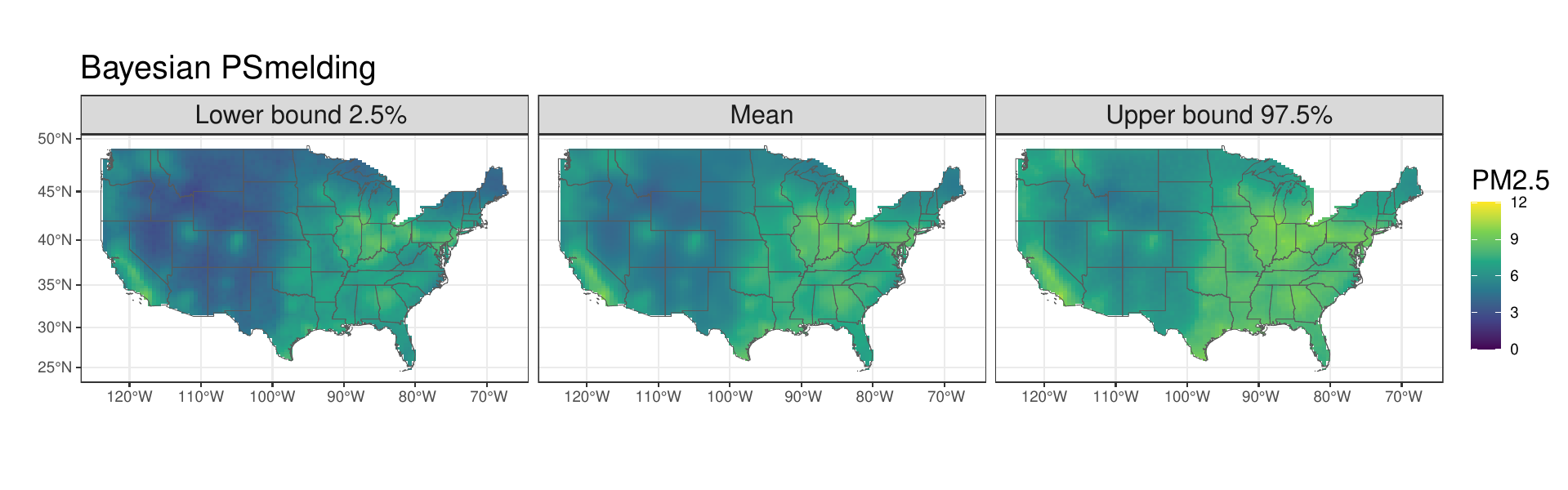}
\label{fig:app_res}
\end{figure}

Maps with the posterior means as well as lower and upper limits of 95\% credible intervals obtained with the proposed melding model adjusting for preferential sampling are shown in Figure \ref{fig:app_res}.
In that figure, we observe that the annual $\text{PM}_{2.5}$ levels are higher in the Eastern US and California than other regions in the USA. The $\text{PM}_{2.5}$ values in northern Utah and southern Wyoming are slightly higher than other regions in central USA. 

\section{Discussion} \label{s:discussion}
In this paper, we proposed a Bayesian hierarchical model to combine spatially misaligned data adjusting for preferential sampling.
This approach specifies a Bayesian melding model that fuses data available at different spatial resolutions by assuming a common Gaussian random field underlying all observations.
The model accounts for preferential sampling by assuming that the Gaussian random field used to model the observations is shared with the log intensity of the point process that originates the locations. A scale parameter is used to control the level of preferentiality.
The model is implemented using the data fusion procedure proposed by \citet{Moraga2017} with the INLA-SPDE approach, which presents computational advantages compared to MCMC methods. 

By means of a simulation study, we show that the model that combines point- and areal-level data accounting for preferential sampling outperforms models that just use one type of data or combine data without accounting for preferential sampling under MSE, MAE, and WD measurements when preferential sampling occurs.
Specifically, we observe that the model that combines data ignoring preferential sampling yields higher MSE, MAE, and WD values, especially when smaller amount of areal data are available. If no preferential sampling exists, PSmelding yields similar error measures as the original melding (true model).
Also, our model presents better performance (according to these error measures) than the model that accounts for preferential sampling but just use point-level data. 

In terms of the parameter estimation, ignoring preferential sampling leads to an unreliable fixed effect posterior mean when areal data are limited, with low coverage probabilities and narrower credible intervals. Our model outperforms other models regarding the obtained posterior mean and credible intervals. As for the hyperparameters, when the spatial dependence is not strong, our model leads to a trade-off between a narrow credible intervals and the ability of covering the true value, whereas the model that just utilizes point-level data tends to have a narrower interval but low CP, and the melding model that does not adjust for preferential sampling presents a wider interval with a higher CP. Comparing the average posterior mean, our model outperforms the other two models.
We also employ our approach to predict $text{PM}_{2.5}$ levels in the USA in 2019. Results show higher $\text{PM}_{2.5}$ values in Eastern US and California.

One limitation of our PSmelding model is that the effect of point-level data may diminish as the number of areas included in the analysis increases. This occurs because the size of the area likelihood tends to dominate, potentially obscuring, the influence of point data. When the number of areas included in the analysis is large, the influence of preferential sampling may become less pronounced or even negligible, thereby limiting the model's ability to accurately account for this phenomenon.

Our proposed model is flexible and can be extended in a number of ways. For example, we could include different structures for the fixed effects, as well as additional random effects to account for different sources of variability.
Inference in such an extended model can still be done by employing the discussed the INLA-SPDE approach.
Also, we may allow the proposed model to account for non-Gaussian non-stationary responses. In this scenario, \citet{cabral2022fitting} and \citet{ngme2} describe how the INLA-SPDE approach can be extended to the non-Gaussian case. 

The presented model was used to combine spatially misaligned data adjusting for preferential sampling. However, it could also be employed to analyze spatio-temporal data available at different spatial and temporal resolutions.
In this setting, we could assume that underlying all observations there is a spatio-temporal Gaussian random field. Then, data aggregated in time, space or space-time could be considered as an average of the Gaussian process at the available resolution. Additionally, the scaling parameter that modulates the preferential degree can be considered fixed or varying in space, time, and space-time \citep{amaral2023model}. 

A special case for spatio-temporal observations would be ``non-temporal preferential sampling'' (e.g., the sampling locations do not change over time). Then, the sampling process $\mathcal{S}$ is temporally independent of the corresponding latent field. In this scenario, we simply treat the temporal point patterns as replicas---i.e., these are realizations from the same point process. This preferential sampling model can be represented as follows 
 \begin{equation}\label{ippp_st}
 \begin{split}
      \mathcal{S} &\sim \text{Poisson Process}(\lambda(\vect{s}, t)) \\
 log(\lambda(\vect{s}, t)) &= \vect{\alpha}
    \vect X(\vect{s}, t) + \gamma(t) \phi(\vect{s}) \\
    \gamma(t) &= \gamma + \epsilon(t),
 \end{split}
 \end{equation}
where $\gamma$ is the common degree of preferential sampling among the replicas and $\epsilon(t)$ is the temporal residual.  

\ruiman{Unlike geostatistical data, when working with areal data, it is typical to have data for every spatial unit. Nonetheless, exceptions may be found. For example, when evaluating the impact of air pollution on population health, it is standard practice to compile health outcomes into aggregated figures for specific areas, such as health districts \citep{wakefield2006health, elliott2007long, watson2021perceptron}, leading to potential issues of preferential sampling in areal data. Our model could be extended to the scenario in which preferential sampling exists in both geostatiscal data and areal data. In that case, the sampling model for point-level data remains the same, i.e., Model \eqref{ippp}, and a Bernoulli model is used to describe the sampling scheme for the areal data. More specifically, let $R_i$ denote the indicator random variable that the areal unit $i$ in Region $\mathcal{D}$ is selected. Also, let $\Phi(B_i) = \frac{1}{B_i} \int_{B_i} \phi(\vect s) d\vect s$ and $\mu(B_i) = \frac{1}{B_i} \int_{B_i} \mu(\vect s) d\vect s$. Then, the areal data with preferential sampling can be described as follows}
\begin{equation}
    \begin{split}
       Y(\vect{B}_i \mid R_i = 1, \Phi(B_i)) &\sim N(\mu(B_i) + \Phi(B_i),\tau_B) \\
       R_i \mid \Phi(B_i) &\sim Bernoulli(p(B_i)) \\
       logit(p(B_i)) &= \vect{\alpha}X(\vect s) + \gamma \Phi(B_i),
    \end{split}
\end{equation}
where all the remaining quantities are defined as before.

In summary, our work allows for the fusion of spatial data available at different resolutions, while also accounting for the possible dependence between the sampling scheme and the underlying process. Our proposed fitting procedure presents computational advantages over MCMC-based methods, as it relies on faster inference techniques. Our model was shown to work well under different (and sometimes, misspecified) scenarios---both for spatial interpolation and parameters estimation. Therefore, although possible to extend, our work presents a first attempt to model spatially misaligned data under the assumption of preferential sampling.

\clearpage

\section{Data availability statement}

The data supporting the findings of this study are described in detail in the main article. The data source information can be found in Section \ref{s:applicaton} of the article.
\section{Funding statement}
The research conducted in this study was supported by GeoHealth research group, King Abdullah University of Science and Technology (KAUST).

\section{Declarations of interest}
The authors declare that they have no conflict of interest.

\bibliographystyle{elsarticle-harv} 
\bibliography{cas-refs}
\pagebreak
\end{document}